\documentstyle[prl,twocolumn,aps,floats,epsfig]{revtex}

\begin{document}
\draft
\flushbottom
\twocolumn[
\hsize\textwidth\columnwidth\hsize\csname @twocolumnfalse\endcsname

\title{ A Finite Element Algorithm for High-Lying Eigenvalues and
Eigenfunctions with Homogeneous Neumann and Dirichlet Boundary Conditions}
\author{G. B\'{a}ez$^{1,3}$, F. Leyvraz$^{1,4}$,
R. A. M\'{e}ndez-S\'{a}nchez$^{1,2,5}$  and  T. H. Seligman$^{1,6}$}
\address{$^1$Centro de Ciencias F\'{\i}sicas,Universidad Nacional Aut\'{o}noma
M\'{e}xico\\
A.P. 48-3, 62250, Cuernavaca, Morelos, MEXICO.}
\address{$^2$Universit\"at G.H. Essen, Fachbereich 7, Physik, 45117, Essen,
Germany.} 
\address{$^3$e-mail:baez@fis.unam.mx $^4$e-mail:leyvraz@fis.unam.mx}
\address{$^5$e-mail:mendez@fis.unam.mx $^6$e-mail:seligman@fis.unam.mx}

\maketitle

\begin{abstract}
We present a finite element algorithm that computes eigenvalues and
eigenfunctions of the Laplace operator for two-dimensional problems with
homogeneous Neumann or Dirichlet boundary conditions or combinations of
either  
for different parts of the boundary. In order to solve the generalized
eigenvalue problem, we use an inverse power plus Gauss-Seidel algorithm. For
Neumann boundary conditions the method is much more efficient than the
equivalent finite difference algorithm. We have cheked the algorithm
comparing the cumulative level density of the espectrum  obtained numerically,
with the theoretical prediction given by the Weyl formula. A systematic
deviation was found. This deviation is due to the discretisation and not to the
algorithm. As an application we calculate the statistical properties of the
eigenvalues of the acoustic Bunimovich stadium and compare them with the
theoretical results given by random matrix theory.\\

\vskip .5 cm

Presentamos un algoritmo de elementos finitos que calcula eigenvalores
y eigenfunciones del operador de Laplace para problemas en dos dimensiones con
condiciones a la frontera de Neumann o Dirichlet o combinaciones de ambas en
distintas partes de la frontera. Para resolver el problema de eigenvalores
generalizado, usamos un algoritmo de potencias inverso m\'{a}s otro de
Gauss-Seidel. Para condiciones a la frontera de Neumann, el m\'{e}todo es
mucho m\'{a}s eficiente que el algoritmo equivalente de diferencias
finitas. Hemos probado el algoritmo comparando la densidad
acumulada de niveles del espectro obtenido
num\'ericamente, con la predicci\'on
te\'orica dada por la f\'{o}rmula de Weyl. Se encontr\'{o} una desviaci\'on
sistem\'{a}tica. Esta desviaci\'on es debida a la discretizaci\'{o}n y no al
algoritmo. Como una aplicaci\'{o}n, calculamos las propiedades
estad\'{\i}sticas de los eigenvalores del estadio de Bunimovich ac\'ustico
y las comparamos con los resultados te\'{o}ricos dados por la
teor\'{\i}a de matrices aleatorias.

\end{abstract}


\vskip .5cm

]


\begin{center}
{\bf I Introduction}
\end{center}

Several years ago Neuberger and Noid\cite{Neu1,Neu2} presented an algorithm
and a FORTRAN program for the successive computation of the high-lying
eigenvalues and eigenfunctions of a time independent Schr\"{o}dinger or
Helmholtz equation. They used an inverse power method with a Gauss-Seidel
procedure for the inversion, and solved the problem with finite differences
on successively finer grids. This program was often used and a
two-dimensional version thereof\cite{Neu3} was adapted for the case of
Laplace operators with homogeneous boundary conditions
\cite{Novaroetal,Mateosetal}. The case of Dirichlet conditions works very well
and requires
minimal adjustments. This is not the case for Neumann conditions.
Computation times increase by orders of magnitude compared to the Dirichlet
case\cite{Novaroetal}; this is not surprising as the treatment of an
irregular boundary, and particularly of corners, is very cumbersome.

The need for such programs arises both in acoustic\cite{BaezG} and
earthquake research\cite{Mateosetal}, as well as for other wave phenomena.
For example, if we want to make statistics of eigenvalues, say for acoustic
systems\cite{BaezG}, we need large numbers of states and therefore efficient
algorithms. In particular, geometries whose high-frequency limit (ray
dynamics) is chaotic, are of great interest. The starting point in this new
field called acoustic chaos is that the time-independent wave equation
(Helmholtz equation) is the same for different systems such as: quantum
billiards\cite{BGS1,Berry}, membranes\cite{Arcosetal} and flat microwave
cavities \cite{Sridhar,Richteretal,Stockmann&Stein}. Thus the statistical
fluctuation measures developed in nuclear physics\cite{Metha,Brodyetal}
have been applied to those systems and to a wide variety of more complicated
systems such as: Chladni's plates\cite{Stein&Stockmann,Legrandetal}, quartz
crystals\cite{Guhr}, aluminum blocks\cite{Guhr2,Weaver}, quantum dots
\cite{Marcusetal}, quantum corrals\cite{Crommieetal,Crommieetal&Heller}, waves
in a ripple tank\cite{Blumeletal}, elastic media with ray
splitting\cite{Ottetal},microwave cavities with ray splitting\cite{Kochetal}
amog others. As rather fine details of the boundary of those systems are
believed to be important, a good representation of the boundary conditions is
essential. Similar arguments will hold if we wish to study the effect of
obstacles inside a cavity or in the old Tenochtitlan lake bed\cite{Baezetal}.

We shall use the finite element method (FEM). It is based on the
minimization of the functional:

\begin{equation}
{\cal F}\left[ \Psi \right] =\int_R\left( \nabla \Psi \right)
^2ds-k^2\int_R\Psi ^2ds, \label {funcional}
\end{equation}
where $R$ is a two dimensional region and $\psi$ is the wave function.

The possibility of solving mixed boundary conditions will be important in
systems with mirror symmetries, in which we may solve 
each non-symmetric part using
mixed boundary conditions. An irregular boundary as well as corners can also 
 easily be
implemented. The FEM\ formalism is based on minimizing
the functional  (\ref{funcional}) not in the complete Hilbert space but only
in a sub-space
spanned by a finite set of piece-wise linear functions, typically defined as
pyramids over hexagonal cells. We call an element the triangles that form
this cell.

The finite difference method becomes very cumbersome for boundary conditions 
involving normal derivatives at irregular boundaries. This is particularly
troublesome for Neumann conditions for which it leads to poor convergence.
Novaro et al. (\cite {Novaroetal}) found in a particular case that computations
would be two orders of magnitude slower for Neumann conditions than for
Dirichlet conditions.

The minimization of the functional (\ref{funcional}) on the 
subspace of Hilbert space and in the non-orthogonal basis 
mentioned above yields the generalized eigenvalue
equation

\begin{equation}
A{\bf x}=\lambda B{\bf x},  \label{generalised}
\end{equation}
with $A_{ij}=\int_R\nabla \psi _i\cdot \nabla \psi _jds$ and
$B_{ij}=\int_R\,\psi _i\psi _jds$, where $\psi _i$ and $\psi _j$ denote the
functions defined around the node $i$ and $j$ of the hexagonal grid
respectively. These functions for simplicity, are taken to be linear

\begin{equation}
\psi _i^k=a_i^kx+b_i^ky+c_i^k  \label{functions}
\end{equation}

with $a_i^k,b_i^k$ and $c_i^k$ constants to be determined for each triangle
$\Delta _k$ of the i-th hexagonal element. The trial functions are then
defined as pyramids of unit height over each hexagonal cell, {\it i.e.} one
function corresponding to each node of the grid. The simplicity of a
piecewise linear basis gives as result a non-orthogonal basis. 
The Neumann boundary conditions are obtained by allowing variations of
the trial functions on the boundary. The Dirichlet boundary conditions can
be obtained by putting the trial functions to zero in the desired part of
the boundary. We refer to the literature for a general and deeper discussion
of the finite element formalism\cite{Strang&Fix,Schwarz,Bethe&Wilson,Mori}.

If the dimension $N$ of the matrices $A$ and $B$ is small enough that they can
be diagonalised, standard techniques for non-orthogonal bases may be used. But
in a typical application, the dimensions are of several thousand to
tenthousands. Yet we
are only interested in a fairly small number of eigenvalues and eigenfunctions
near the low end of the spectrum. We thus have to use some method that makes
explicit use of the sparseness of the matrices $A$ and $B$. We shall see in
the next section, that a combination of 
the inverse power and Gauss-Seidel methods proposed by
Neuberger and Noid\cite{Neu1,Neu2} can be generalised to solve Eq.
(\ref{generalised}).

Thus in the following section we show how the inverse power method can be
applied when a non-orthogonal basis is used.  Next we discuss 
how to implement this for finite differences as well as a number of
tricks that can be used to accelerate the numerical procedure and comment on
the performance of the program. In section IV we apply the program to the
acoustic stadium, analyze the resulting spectra and the corresponding
states. We give a correction to the spectral density based on an analysis of
the equations infinite elements for the exactly solvable rectangle discussed in the appendix, Finally we present some conclusions.

\begin{center}
{\bf II The inverse power method in a non-orthogonal basis}
\end{center}

We shall transform the Eq.\ (\ref
{generalised}) by left multiplication with $B^{-1}$ to the form
\begin{equation}
B^{-1}A{\bf x}=\lambda {\bf x}
\end{equation}
As usual a power $(B^{-1}A)^{-M}\equiv (A^{-1}B)^M$ applied to an arbitrary
initial vector ${\bf x}^0$ will successively select the lowest
eigenvector corresponding to eigenvalue $\lambda _1$ as it will appear with
a power $(1/\lambda _1)^M$.

The problem now seems to be that $A^{-1}$ is no longer a sparse matrix, but
this can be averted in the procedure of applying $A^{-1}B$. Thus we need to
know the product
\begin{equation}
{\bf y}\equiv A^{-1}B{\bf x}^0.  \label{defin}
\end{equation}

In order to obtain this product we define $\widetilde{ {\bf x}}^0\equiv B%
{\bf x}^0$ giving for Eq. (\ref{defin})
\begin{equation}
{\bf y}=A^{-1}{\bf \widetilde{x}}^0
\end{equation}
Multiplying by A by the left we obtain

\begin{equation}
A{\bf y}=\widetilde{{\bf x}}^0
\end{equation}
that can be solved alternatively by Jacobi or Gauss-Seidel procedures which
will work well as all matrices involved continue to be sparse. Summarising:
the Gauss-Seidel Method can be utilized because we do not need the inverse
matrix $A^{-1}$ but the product $A^{-1}\widetilde{{\bf x}}^0$.

Up to here we
have only specified how to obtain the lowest state; for excited states
the usual procedure is to orthogonalize the space in which we carry out the
variation on all states that have already been calculated. Here again the
non-orthogonality of our basis has to be taken into account; indeed, the
eigenstates of the Laplace operator are orthogonal and we have to derive the
consequences this has for eigenvectors in our non-orthogonal basis. If $\Psi
_i$ and $\Psi _j$ are eigenstates of the Laplace operator and are expanded
as with coefficients $\alpha _m^l\,;\,l=i,j$ that form vectors ${\bf x}_l$
we can readily check that
\begin{equation}
\int_R\Psi _iB\Psi _jds=\delta _{ij}  \label{orto}
\end{equation}
implies
\begin{equation}
{\bf x}_i^t B{\bf x}_j=\delta _{ij}  \label{Borto}
\end{equation}
and vice-versa. Thus we replace the usual orthogonalization by what we may
call a $B$-orthogonalization, i.e. we require the new vectors to be
orthogonal on $B{\bf x}_j$ (where $j$ indicates the calculated
eigenstates) thus guaranteeing the validity of Eq. (\ref{Borto}) and by
consequence Eq. (\ref{orto}).

\begin{center}
{\bf III Implementation of finite elements and convergence}
\end{center}

Once we know the grid and the matrix elements of $A$ and $B$ both of which
will be evaluated in the end of this section, we are in principle ready to
write our program. As usual a program consists in part of an efficient
algorithm which we have presented, and in part of a bag of semi-empirical
tricks that tend to repeat themselves in different guises again and again.
The efficiency of the latter is quite problem-dependent and the
corresponding parameters must be adjusted to optimize operation of the
program in every case. We shall give some recommendations that ought to work
reasonably, but we urge the user to fiddle around with these parameters.

When using Neumann boundary conditions, the lowest eigenvalue is zero and
its corresponding eigenfunction is a constant. Yet the Gauss-Seidel inversion
requires
positive definite eigenvalues. This is obtained by adding an arbitrary
constant $C$ multiplied by $B$ to the Laplace operator. If we choose this
constant large we will need few Gauss-Seidel iterations as the operator is
near diagonal. On the other hand, we will loose precision and convergence
speed in the inverse power process as neighbouring eigenvalues will have
their inverse very close to each other. A good balance seems to be to choose
the constant smaller than, but of the order of, the largest eigenvalue which
we want to obtain. In the case of Dirichlet conditions this parameter can
also be introduced as a means of improving convergence exclusively. Adequate
choices of this parameter may improve computation time by a factor of $\sim
2 $.

Superconvergence, or over-relaxation, is another standard tool to improve
convergence. We introduce a factor $1<\alpha <2$ and at each step of any
iteration from ${\bf x}_n\to {\bf x}_{n+1}$ we replace ${\bf x}
_{n+1}$ by ${\bf x}_n+\alpha ({\bf x}_{n+1}-{\bf x}_n)$ thus
correcting a little more than the iteration warrants. For the Gauss-Seidel
iterations values of $\alpha \approx 1.5$ yield improvements in computing
time of the order of $\sim 2$, which is consistent with what we found for
finite differences. But a careful analysis for the stadium shows that at
least in this case for a quite precise value of $\alpha =1.75$ we find an
acceleration by a factor of $\sim 3$. On the other hand, in the power
iteration, improvements are not very significant and only values of $\alpha $
near 1 seem acceptable. We do not use superconvergence in this context, but
again we must warn that it might be very significant in other cases. All
these parameters were carefully explored by M\'{e}ndez\cite{MendezR}.

Another option is the stepwise introduction of finer grids, with
interpolated results from the rougher grid results as starting point. This
idea was extensively exploited in the finite difference programs of
Neuberger and Noid\cite{Neu1,Neu2} and gave excellent results both
shortening the computation time by giving a good trial function on the
finest grid and allowing a considerable improvement of the eigenvalue upon
extrapolation. Unfortunately these advantages diminish even in their case as
we go to higher states, for which only the finest grid is acceptable (finer
ones would yield too large matrices). For this reason we have not
implemented these procedures for finite elements at this point.

Now we return with the problem of establishing the grid that defines our
finite elements and of calculating the matrix elements of $A$ and $B$. For
this we use a standard method\cite{Strang&Fix,Schwarz,Bethe&Wilson,Mori}
summarised in the following steps:

\begin{enumerate}
\item  We immerse the region $R$ in a quadratic grid. The triangles of the %
elements are defined using the sides of the squares and in addition one of %
the diagonals.

\item  We redefine the grid and triangles along the boundary by considering %
all points that lie just outside our contour. We then move the exterior %
points along the edges of the squares or the selected diagonal, to the %
boundary, so as not to change the topology of the grid and its connections.
\end{enumerate}

The corresponding integrals are evaluated using a
change of variable with a linear transformation. The transformation can be
found solving a linear system of 3 equations. Evaluating the constants of
Eq.~(\ref{functions}) and the Jacobian of the transformation (in this case
the one half area of the triangle, because the transformation is
linear) we obtain for the integrals:

\begin{equation}
A_{ij}=\sum_kS(\Delta _k)(a_i^ka_j^k+b_i^kb_j^k)
\end{equation}
and
\begin{equation}
B_{ij}=\left\{ \matrix{\ \sum_k2S(\Delta _k)\frac 1{24},i\ne j,\cr\cr
\sum_k2S(\Delta _k)\frac 1{12},i=j}\right.
\end{equation}
Here $S(\Delta _k)$ is the area of triangle $\Delta _k$ and the sum is made
over the number of triangles common to both elements $i$ and $j$.

The routines consist of two main parts. The first computes all elements
different from zero. The elements are put in two matrices of $N\times 7$ to
use the sparseness of matrices. In order to retain the original positions of
each element we construct an index matrix. The second part solves the
generalized system of equations using the inverse power iteration or a
standard diagonalization freeware routine \cite{http1}, if the
dimensions of the matrices are small (less than $\approx 5000\times 5000$).
We obtain $\approx 500$ eigenvalues with error less than $5\%$ of the
average spacing between levels.

The input for the first routine is a set of points localised on the border
of the region of interest and a list of intervals of this
enumeration in which we want Dirichlet boundary conditions. We assume
Neumann boundary conditions in the rest of the border.

For the second routine the inputs are the
number of eigenvectors, the tolerance for the Gauss Seidel and inverse power
iterations, the constant $C$, and the superconvergence factor $\alpha $. The
output consists of two files with the corresponding eigenvalues and
eigenvectors. 

\begin{center}
{\bf IV Application to the acoustic stadium}
\end{center}
By way of example we apply the program to some particular two-dimensional
region $R$; such as the Bunimovich stadium which is completely chaotic in
classical mechanics. This geometry consists of two
semicircles of radius $r$, joined by two straight lines with length $2l$
\cite{Bunimovich}. We will apply pure Neumann boundary conditions. On
a grid of $\sim 3000$ points we obtain, by the inverse
power iteration method described above, $200$
eigenvectors and eigenvalues with a CPU time of approximately
$200$ $\sec $. on an ALPHA Work-Station.

To analised a discrete sequence of eigenvalues we first define the cumulative
level density or staircase function
\begin{equation}
N(E)=\sum_{i=1}^N\Theta (E-E_i)  \label{nde}
\end{equation}
and its derivative $\rho (E)$ (the level density),

\begin{equation}
\rho (E)=\sum_{i=1}^N\delta (E-E_i).
\end{equation}
Here $\Theta $ and $\delta $ are the Heaviside and Dirac delta functions
respectively. The staircase function is usually divided in a smooth part
plus a fluctuating part:

\begin{equation}
N(E)=\overline{N}(E)+N_{fluct}(E).
\end{equation}

\begin{figure}[tbp]
{\hspace*{-0.5cm}\psfig{figure=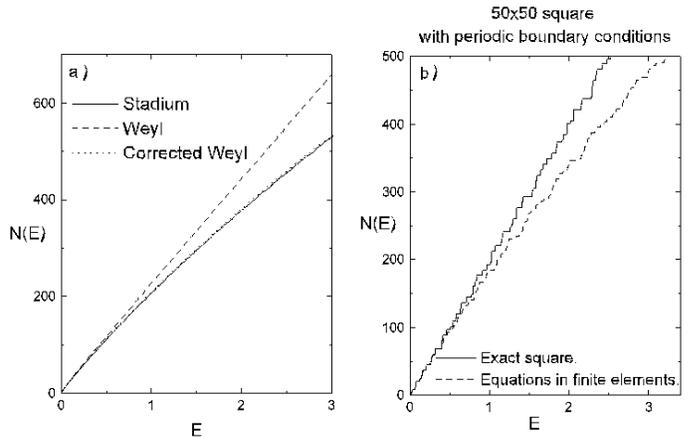,height=6cm}} \vspace*{.12in}
\caption{a) Cummulative level density for a quarter of
Bunimovich stadium with Neumann boundary conditions and discretised with
2791 elements. The dashed curve correspond to the theoretical result given
by the Weyl formula and the dotted curve is the Weyl formula corrected by
the polynomial from the equations in finite elements for the square. b)
Cummulative level density for a $50\times 50$ square with periodic boundary
conditions. As reference the theoretical prediction for the square is also
plotted.} \label {fig:1}
\end{figure}
   
The cumulative level density obtained by the finite element method for the
stadium with Neumann boundary conditions is plotted in Fig. 1a. For
comparison the average level density

\begin{equation}
\overline{N}_{Weyl}(E)=(AE+P\sqrt{E}+\kappa ) \,
\end{equation}
obtained from the Weyl formula \cite{Gutzwiller} is depicted. Here $A$ is the
area of the billiard, $P$ is the length of its perimeter and $\kappa $ is a
constant that contains information on the topological nature of the billiard
and the curvature of its boundary. From this figure we can see that the finite
element method gives a systematic deviation of the eigenvalues\cite{BaezG}.
This systematic deviation in the cumulative level density is due to the
discretisation of the finite element. To show this we calculate in the
appendix the eigenvalues $\lambda _{n,m}$ for the equations in finite
elements for an $a\times a$ square with periodic boundary conditions. The
final equation is

\begin{equation}
\lambda _{n,m}=\frac{4-2\left( \cos \left( k_x\right) +\cos \left(
k_y\right) \right) }{\frac 12+\frac 16\left( \cos \left( k_x\right) +\cos
\left( k_y\right) +\cos \left( k_x+k_y\right) \right) }\,  \label{eqsinfe}
\end{equation}
where $k_x=\frac{n\pi }a$ and $k_y=\frac{m\pi }a$ are the wave number on the
$x$ and $y$ directions, $a$ is the size of the side and $n,m=0,\pm 1,\pm
2,\ldots $

In Fig. 1b we show the cumulative level density obtained from Eq. (\ref
{eqsinfe}) for a $50\times 50$ square. We also plotted the cumulative level
density for the exact square ($\propto k_x^2+k_y^2$). The one coming from
the diagonalisation is indistinguishable from the obtained by the Eq. (\ref
{eqsinfe}). The systematic deviation observed in the square with periodic
boundary conditions will be used to correct the Weyl formula for
arbirary-shaped billiards and arbitray boundary conditions. To do this we
calculate the difference between the fits for both cumulative level
densities --Eq. (\ref{eqsinfe})-- and the exact square. The polynomial for the
difference is $A(4.60055-13.75335E-10.44418E^2+0.45601E^3)/2500$, where $A$
is the area of the billiard.
If this polynomial is added to the Weyl prediction,
the resulting curve yields good agreement
up to $\sim 400$ eigenvalues in the stadium (See Fig. 1a).

We shall study the fluctuation properties of the spectrum. In order to do
this for a typical sequence of eigenvalues, it is necessary to suppress the
secular variation. This ``unfolding'' of the levels is done through the
mapping

\begin{equation}
E_i\longmapsto E_i^{\prime }=\overline{N}(E_i).
\end{equation}

\begin{figure}[tbp]
{\hspace*{-.5cm}\psfig{figure=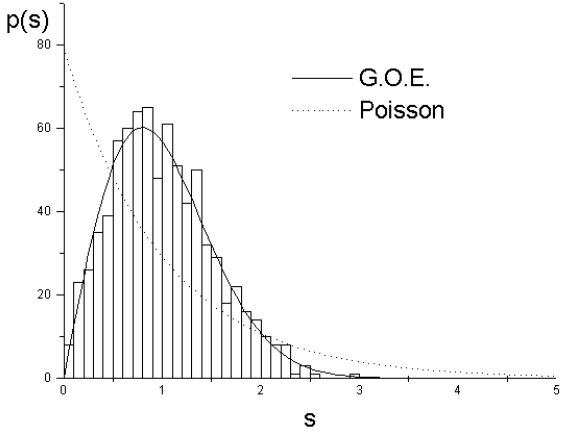,height=6cm}} \vspace*{.12in}
\caption{Nearest-Neighbour spacing distribution for the
GOE (solid line), and for the Poisson sequence (dotted line). The one
obtained for the stadium (200 eigenvalues for each symmetry--NN,DD,ND,DN--
of a quarter of the Bunimovich stadium with Neumann boundary conditions) is
depicted by the histogram.} \label {fig:2}
\end{figure}

The effect of such mapping on the original sequence is that the new sequence
has on the average a constant spacing equal to one. Although we should use the
Weyl formula corrected by the polynomial, we can use a polynomial fit
$\overline{N}(E)$ for $N(E)$ that takes into account the systematic deviation
due to the discretisation. We can then calculate different statistical
measures of the fluctuating part of the spectrum. The first statistic we
shall use is the nearest-neighbour spacing distribution $p(s)$ where
$s=E_{i+1}^{\prime }-E_i^{\prime },$ which gives information on the short
range correlations. The $p(s)$ for the stadium is given in Fig. 2. Notice
that it agrees with the values predicted by the gaussian orthogonal ensemble
(GOE) typical for chaotic systems. For completeness, the Poisson
case, typical for integrable systems, is also depicted. The spectrum of the
acoustic Bunimovich stadium shows $p(s\rightarrow 0)\rightarrow 0$. This
behaviour is called level repulsion. In fact the nearest-neighbour spacing
distribution agrees with

\begin{equation}
p_{Wigner}(s)=\frac \pi 2s\exp (-\frac \pi 4s^2),\qquad s\geq 0 \,  \label{pds}
\end{equation}
know in spectral statistics as the Wigner surmise, which is very close to
the GOE\ prediction.

We can also define the $k$th-neighbour spacing distribution $p(k;s_k).$ Now
$s_k=E_{i+k+1}^{\prime }-E_i^{\prime }$, so that $p(s)=p(0;s_0\equiv s)$. It
is well known \cite{Brodyetal} that these distributions tend to a normal
distribution as $k$ grows. Since $\left\langle s_k\right\rangle =k+1$, the
only relevant parameter left is the width $\sigma (k)$ of the distribution.
In Fig. 3 we show $\sigma (k)$ for the stadium, GOE and Poisson cases.

\begin{figure}[tbp]
{\hspace*{-.5cm}\psfig{figure=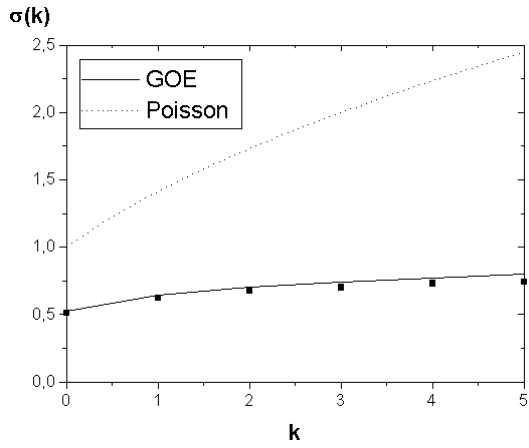,height=6cm}} \vspace*{.12in}
\caption{The width $\sigma (k)$ of the $k$th-neighbour
spacing distribution $p(k;s_k)$ as a function of $k$ for the GOE (solid
line), the Poisson (dotted line) and the acoustic stadium (squares).} \label {fig:3}
\end{figure}

The correlation coefficient between adjacent spacings is another short range
fluctuation measure. For the acoustic stadium we obtained $C=-0.26$ near to
the GOE value ($C_{GOE}=-0.27$) and far from the Poisson value
($C_{Poisson}=0$).  

Another commonly used statistic is the number variance $\Sigma^2(L)$, defined
as the second moment of the number of levels $\nu (L)$ within an interval of
length $L$, and given for the stadium, GOE and the Poisson cases in Fig. 4.
For the stadium and GOE cases and for large $L$, $\Sigma^2(L)\approx \ln (L)$
indicating a very rigid sequence.

\begin{figure}[tbp]
{\hspace*{-.5cm}\psfig{figure=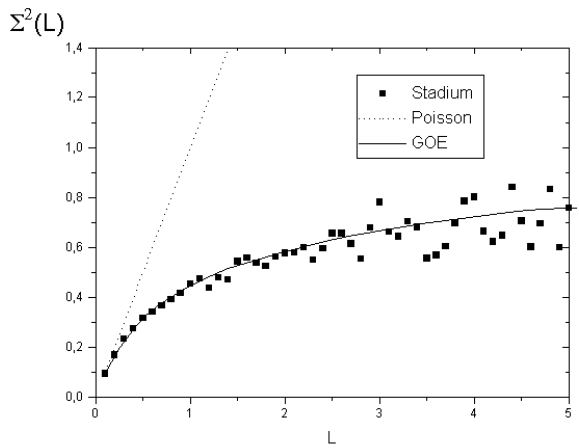,height=6cm}} \vspace*{.12in}
\caption{The number variance $\Sigma^2(L)$ as a function of the length $L$ of
the interval for the same cases as in Fig. 3.} \label {fig:4}
\end{figure}

The number variance depends exclusively on the two-point function. We
consider further moments
\begin{equation}
 \Sigma^k(L)=\left\langle \left[ \nu (L)-\left\langle \nu
(L)\right\rangle \right] ^k\right\rangle
\end{equation}
that depend on higher correlations. To emphasise the three- or four-point
properties\cite{BHP}, it is useful to consider the skewness

\begin{equation}
\gamma _1(L)=\Sigma^3(L)\times \left[ \Sigma^2(L)\right]
^{-3/2}  \label{gamma1}
\end{equation}
and the excess
\begin{equation}
\gamma _2(rL)=\Sigma^4(L)\times \left[ \Sigma^2(L)\right]
^{-2}-3.  \label{gamma2}
\end{equation}

The numerical values obtained for the stadium are shown in Figs. 5 and 6.

\begin{figure}[tbp]
{\hspace*{-.5cm}\psfig{figure=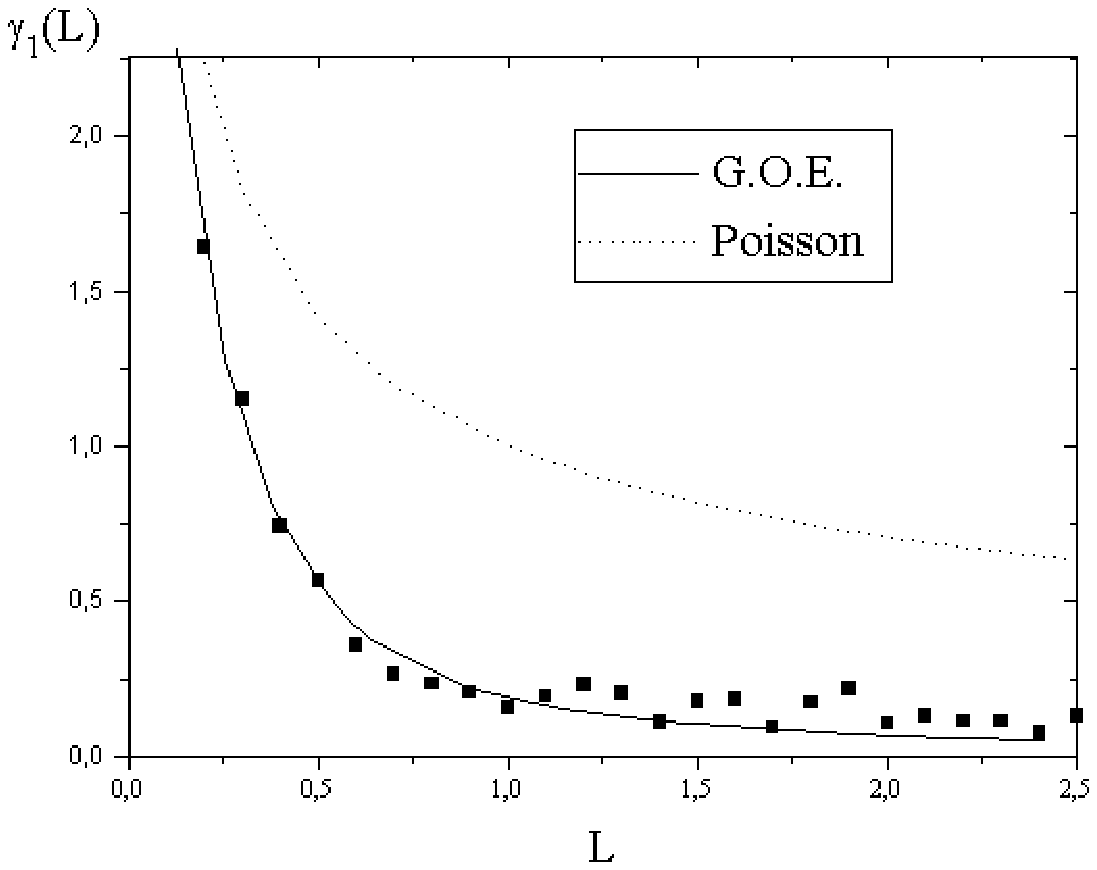,height=6cm}} \vspace*{.12in}
\caption{The skewness $\gamma _1(L)$ --Eq. (20)-- as a function of
the length $L$ of the interval for the same cases as in Fig. 3.} \label{fig:5}
\end{figure}

\begin{figure}[tbp]
{\hspace*{-.5cm}\psfig{figure=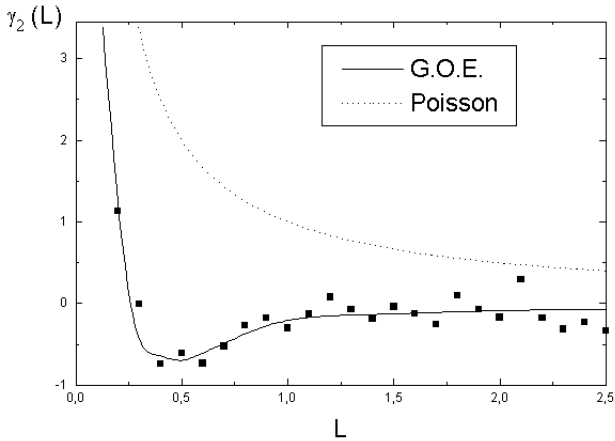,height=6cm}} \vspace*{.12in}
\caption{The excess $\gamma _2(L)$ --Eq. (21)-- as a function of the
length $L$ of the interval for the same cases as in Fig. 3.}
\label {fig:6}
\end{figure}

In many instances the Fourier transform of the spectra has also proven
useful. It contains the same information as the spectrum itself. On the
other hand the power spectrum:

\begin{equation}
\left| C(t)\right| ^2=\frac 1{2\pi }\left| \int\limits_{-\infty }^\infty
e^{-2i\pi Et}\rho (E)dE\right| ^2,  \label{pow}
\end{equation}
depends exclusively on the two-point function\cite{Lombardietal}. The
short-range part of $P(t)$ gives specific information concerning the
long-range stiffness. Numerical values are given in Fig. 7.

\begin{figure}[tbp]
{\hspace*{-.5cm}\psfig{figure=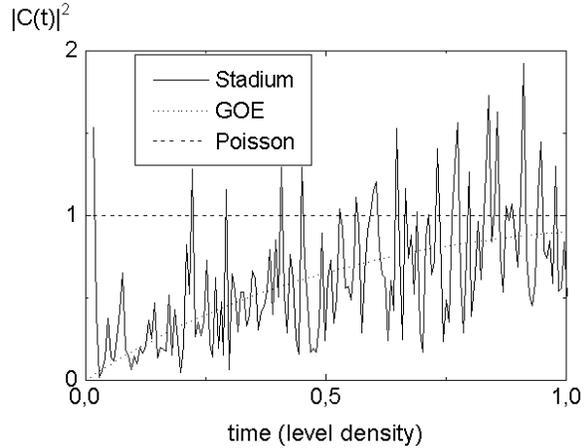,height=6cm}} \vspace*{.12in}
\caption{The power spectrum $\left| C(t)\right| ^2$ --Eq. (22)-- as a
function of the dimensionless variable $t$, for the same cases as in Fig. 3.
We eliminated the divergence at the origin due to the finite range of the
spectrum.} \label {fig:7}
\end{figure}

\begin{figure}[tbp]
{\hspace*{-.5cm}\psfig{figure=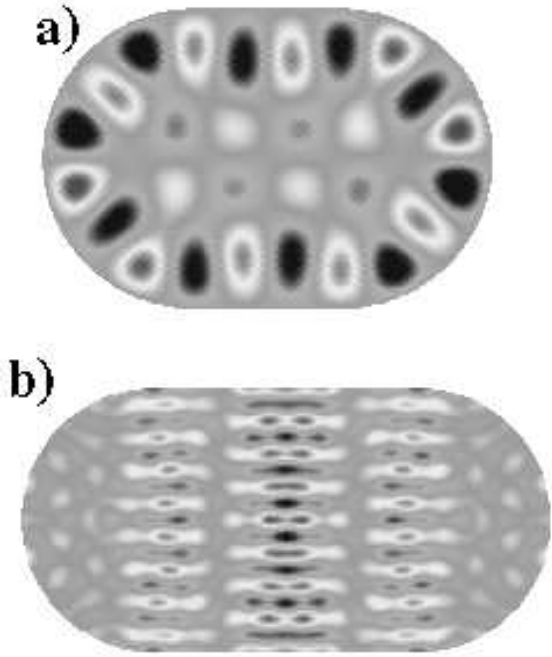,height=9cm}} \vspace*{.12in}
\caption{Eigenfunctions of Bunimovich stadium calculated by finite element
method for a quadrant and reflecting with respect to both axis: a) Dirichlet
boundary conditions with a relation $l/r = 1/2$. This figure show a
whispering gallery state; b) Neumann boundary conditions with
$l/r=1$. Typical state scarred by a near ``bouncing-ball'' orbit.}
\label{fig:8} 
\end{figure}

The eigenfunctions for the rectangle were also successfully compared with
the exact solution. In Fig. 8 we show two eigenfunctions of the stadium: one
of them with Dirichlet boundary conditions which shows typical feature of the
whispering gallery states\cite{McDonald&Kauffman}, and the last one, with
Neumann boundary conditions shows a near bouncing ball state. There are other
kinds of features in different eigenfunctions, like scars
\cite{Heller1,Heller2}, and others that resemble noise \cite{Berry1,Berry2}.
Some which have been reproduced by the autors with this algorithm can be seen
in ref. \cite{Arcosetal}.

Finally, we performed all the calculation on a quarter stadium and
used Neumann (N) or Dirichlet (D) boundary conditions on the symmetry
lines. This implies that we studied the symmetric or antisymmetric solutions
with respect to both reflection symmetries of the stadium. The full solution,
shown in the figures, is recovered making the corresponding reflections respect
each symmetry axis.

\begin{center}
{\bf V Conclusions}
\end{center}

We have presented an algorithm based on the finite element method that
computes eigenfunctions and eigenvalues of the two-dimensional Helmholtz
equation with mixed Neumann and Dirichlet boundary conditions. The algorithm
is divided in two parts: one that computes the matrix elements and another
that diagonalizes the generalised eigenvalue equation using the inverse
power and Gauss-Seidel Methods. The Gauss-Seidel method runs efficiently if
we use over-relaxation where the gain in computing time peaks at a factor of
$\approx 3$ around the value 1.75 for the superconvergence
(over-relaxation). A systematic error in frequencies was found. This error
is due to the discretisation and can be estimated by using the eigenvalues
of the equations in finite elements. The programs were used to calculate the
eigenvalues of the acoustic stadium. The fluctuation measures of the
eigenvalues were compared with the random matrix predictions.

The algorithm is very useful in diverse branches of wave physics. The
program can obtain eigenfunctions and eigenvalues of two-dimensional
acoustic cavities, two-dimensional microwave cavities, membranes, quantum
billiards, elastic valleys in certain approximations, etc., and can be
readily generalised to other problems.

\begin{center}
{\bf Acknowledgements}
\end{center}

This work was supported by the UNAM---CRAY Research Inc. project SC101094,
by UNAM-DGAPA project IN106894. G. B\'{a}ez and R. A. M\'{e}ndez received
fellowships by UNAM-DGAPA. We want to thank to the IF-UNAM in which, the main
part of this work was developped.

\newpage
\begin{center}
{\bf Appendix: Eigenvalues for the equations in finite elements.}
\end{center}

In this appendix we deduce the equations in finite elements for a square
with periodic boundary conditions. If we denote by

\begin{equation}
\text{\bf x}=e^{i(k_xn+k_ym)}  \label{periodic}
\end{equation}
the amplitude in the grid point $(n,m)$, the Eq. (\ref{generalised}) for the
bulk element is given by
\begin{eqnarray}
&&(4 -e^{ik_x} -e^{ik_y} -e^{-ik_x}
-e^{-ik_y})\text{\bf x} \nonumber \\
&&= \lambda _{n,m} \Big( \frac 12 +\frac 1{12}e^{ik_x} +
\frac 1{12}e^{ik_y} +\frac 1{12}e^{-ik_x} +
\frac 1{12}e^{-ik_y} \\
&& + \frac 1{12}e^{i(k_x+k_y)}
 +\frac 1{12}e^{-i(k_x+k_y)} \Big) \text{\bf x.}  \nonumber
\end{eqnarray}
Here $k_x=\frac{n\pi }a$ and $k_y=\frac{m\pi }a$. We also assumed that the
size of the grid is one and that for the bulk $A_{i,i}=4,$ $A_{i,j\pm
1}=A_{i\pm 1,j}=-1$, $B_{i,j\pm 1}=B_{i\pm1,j}=B_{i\pm 1,j\pm 1}=\frac 1{12}$
and $B_{i,i}=\frac 12$. The final form for the eigenvalue
equation in finite elements is given by

\begin{eqnarray}
&&4-2\left( \cos (k_x)+\cos (k_y)\right) \nonumber \\
&&=\lambda _{n,m}\left( \frac 12+\frac 16\left( \cos (k_x)+\cos (k_y)+
\cos (k_x+k_y)\right) \right) .
\end{eqnarray}

\end{document}